\documentclass[twocolumn,showpacs,preprintnumbers,amsmath,amssymb,prl]{revtex4}

\usepackage[dvips]{graphicx}
\usepackage{dcolumn}
\usepackage{amssymb}
\usepackage{amsfonts}
\usepackage{bm}
\usepackage{epstopdf}
\begin{document}

\preprint{APS}
\title{Decay of a discrete state resonantly coupled to a continuum
of finite width}
\author{E. Kogan}
\email{Eugene.Kogan@biu.ac.il}
\affiliation{Department of Physics, Bar Ilan University, Ramat Gan 52900,
Israel}
\affiliation{Max-Planck-Institut fur Physik komplexer Systeme,  Dresden 01187, Germany}
\date{\today}

\begin{abstract}
A simple quantum mechanical model consisting of a discrete level
resonantly
coupled to a continuum of finite width, where the coupling can be
varied from perturbative to strong (Fano-Anderson model), is considered.
The particle is
initially localized at the discrete level, and the time dependence
of the amplitude to find the particle at the discrete level is
calculated   without resorting to perturbation theory.
The deviations from the exponential decay
law, predicted by the Fermi's Golden Rule, are discussed.
We also study analytic structure of the Green's function (GF) for the
model. We analyze the GF poles,
branch points and Riemann surface, and show how the  Fermi's Golden
Rule, valid in perturbative regime for not to large time, appears in
this context. The knowledge of analytic structure of the GF in
frequency representation opens opportunities for obtaining easy for numerical calculations formulas
for the GF in time representation, alternative to those
using the spectral density.

\end{abstract}
\maketitle

\section{Introduction}

The transition of a quantum particle from an initial discrete state
of energy $\epsilon$ into continuum of final states  is considered
in any textbook on quantum mechanics. It is well known that
perturbation theory approach, when  used to solve the problem, leads
to Fermi's Golden Rule (FGR), which predicts the exponential decrease of
the probability to find the particle in the discrete state. It is
also well known that, even for a weak  coupling between the discreet
state and the continuum, this result (exponential decrease of
probability) has a finite range of applicability, and is not valid
either for very small or for very large time (see e.g.
Cohen-Tannoudji et. al. \cite{cohen}). This complies with the
theorem proved 50 years ago, and stating that for quantum system
whose energy is bounded from below, i.e., $(0,\infty)$ the
exponential decay law cannot hold in the full time interval
\cite{khalfin,fonda,nakazato}. The same statement remains valid
when the discrete state is coupled to the continuum
bounded both from below and above, the model we presently consider.

Formally speaking, the model we consider
is exactly solvable (and was solved a long time ago).  The solution in the frequency representation obtained within the
 Green's functions (GF) formalism is presented in \cite{mahan}, where the model
is called the Fano-Anderson model \cite{fano}). On the other hand, the qualitative behavior of the integral which represents the solution
is far from being obvious. Also, in some cases this integral is not convenient for numerical calculations.
We posted two connected papers on the subject in the arXive in 2006 \cite{kogan1} and published part of the results in 2008 \cite{kogan2}.
After that we laid to rest the activities for more than 10 years. However, quite recently we realised that the
 attention of the community to the aspects of the theory mentioned above  was again attracted,
 and our posters were cited. This occurred
in the papers studying impurity coupled to a lattice with disorder \cite{visuri}, non-ergodic delocalized states for the population transfer\cite{alt1}, and similar states in quantum spin glass \cite{alt2}.
this is why we decided to return to our previous results, combine and edit them and make them accessible to broad audience.

In this paper we  concentrate on the calculation of the  time dependent non-decay amplitude,
the stage which is typically not given proper attention to
within the GF formalism \cite{mahan}.
We analyze the relation between the exact results and  those given by the FGR.

In this paper we would also like to see the
Green's function of the problem in a broader context, as a multi valued function,
and study its analytical
structure. On simple examples  we'll
study  that Green's
function (in frequency representation)  branch points, poles and Riemann
surface. This study  prompts effective algorithms to numerically calculate the Green's function.
It allows also to connect between the
frequency representation of the Green's function (used in
calculation of the spectral line intensity) and the time
representation.

\section{Decay}

 Our system consists of the continuum
band, the states bearing index $k$, and the discrete state $d$,
having energy $\epsilon$. The Hamiltonian of the problem is
\begin{eqnarray}
\label{ham}
H=\sum_k\omega_k\left|k\right>\left<k\right|
+\epsilon \left|d\right>\left<d\right|
+\sum_k\big(V_k\left|k\right>\left<d\right|
+h.c.\big),
\end{eqnarray}
where $\left|k\right>$ is a band state and $\left|d\right>$ is the
state localized at site $d$; h.c. stands for the Hermitian
conjugate. The wave-function  can be presented as
\begin{eqnarray}
\label{ab}
\psi(t)=g(t)\left|d\right>+\sum_{k} b(k,t)\left|k\right>,
\end{eqnarray}
with the initial conditions $g(0)=1$, $b(k,0)=0$. Notice that the
non-tunneling amplitude is just the appropriate GF in time
representation. Schroedinger Equation for the model considered takes
the form
\begin{eqnarray}
\label{dif}
i\frac{dg(t)}{dt}=\epsilon g(t)+\sum_{k}V_k^* b(k,t)\nonumber\\
i\frac{db(k,t)}{dt}=\omega_k b(k,t)+V_k g(t)
\end{eqnarray}

Making Fourier transformation (Im $\omega>0$)
\begin{eqnarray}
g(\omega)=\int_{0}^{\infty}g(t)e^{i\omega t}dt,
\end{eqnarray}
we obtain
\begin{eqnarray}
\label{eq}
-i+\omega g(\omega)&=&\epsilon g(\omega)+\sum_{k}V_k^* b(k,\omega)\nonumber\\
\omega b(k,\omega)&=&\omega_k b(k,\omega)+V_k g(\omega).
\end{eqnarray}
For the amplitude to find electron at the  discrete level,
straightforward algebra gives
\begin{eqnarray}
\label{int}
g(t)=\frac{1}{2\pi i}\int g(\omega)e^{-i\omega t}d\omega,
\end{eqnarray}
where
\begin{eqnarray}
\label{exact}
g(\omega)=\frac{1}{\omega-\epsilon-\Sigma(\omega)},\;
\end{eqnarray}
and
\begin{eqnarray}
\label{sigma}
\Sigma(\omega)=\sum_k\frac{|V_k|^2}{\omega-\omega_k}
\end{eqnarray}
The integration in Eq. (\ref{int}) is along any infinite straight
line parallel to real axis in the upper   half plane of the complex
$\omega$ plane. Notice that $g(\omega)$ is the  GF in frequency
representation. The quantity $\Sigma(\omega)$  is self-energy (or
mass operator).

For tunneling into continuum, the sum in Eq. (\ref{sigma})
should be considered as an integral, and Eq. (\ref{sigma}) takes the form
\begin{eqnarray}
\label{sigma3}
\Sigma(\omega)=\int_{E_b}^{E_t}
\frac{\Delta(E)}{\omega-E}dE,
\end{eqnarray}
where
\begin{eqnarray}
\Delta(E)=\sum_{k}
|V_k|^2\delta(E-\omega_k),
\end{eqnarray}
where and the limit of integration are the band bottom $E_b$ and the
top of the band $E_t$. We would like to calculate integral
(\ref{int}) closing the integration contour  by a semi-circle of an
infinite radius in the lower half-plane. Thus we need to continue
analytically the function $g(\omega)$ which was defined initially in
the upper half plane (excluding real axis) to the whole complex
plane. We can do it quite simply, by considering Eqs. (\ref{exact})
and (\ref{sigma3}) as defining propagator
 in the whole complex plane, save an interval of
real axis between the points $E_b$ and $E_t$, where Eq.
(\ref{sigma3}) is undetermined. (Propagator analytically continued
in such a way we'll call the standard propagator.) Thus the integral
is determined by the integral of the sides of the branch cut between
the points $E_b$ and $E_t$.
\begin{figure}
\includegraphics[angle=0,width=0.45\textwidth]{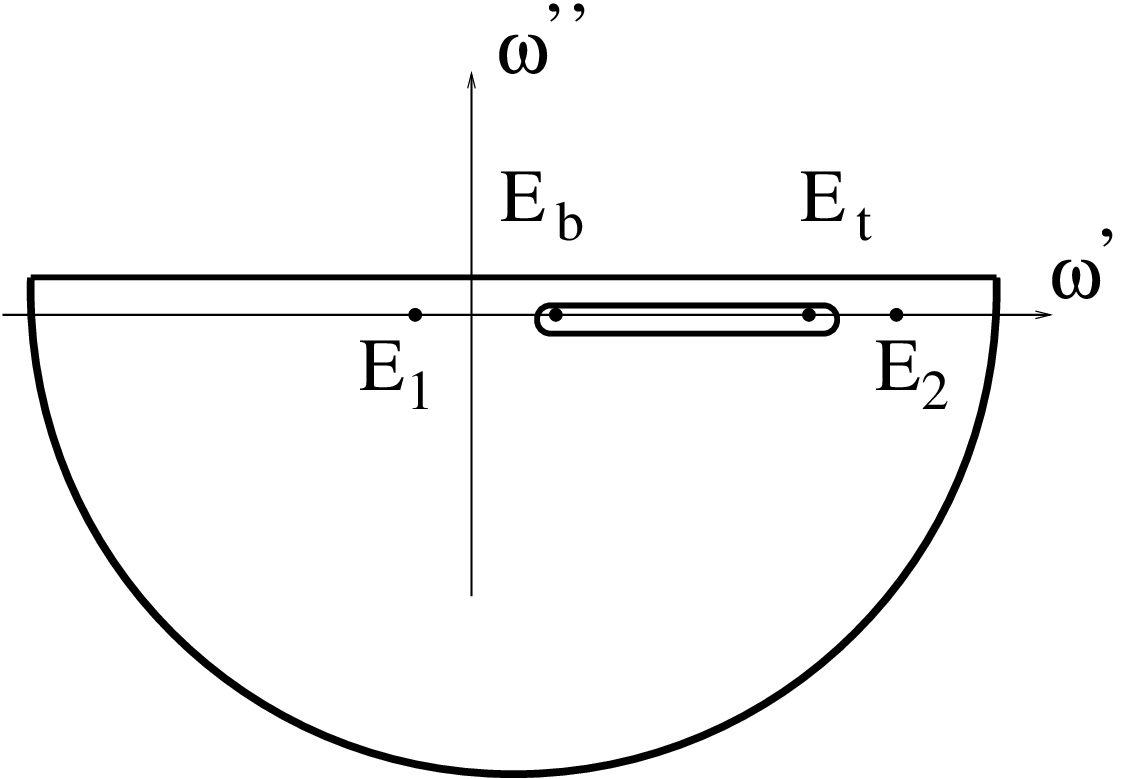}
\caption{\label{fig:cont} Contour used to evaluate integral
(\ref{int}). Radius of the arc goes to infinity.}
\label{contour}
\end{figure}
The real part of the self-energy $\Sigma'$ is continuous across the
cut, and the imaginary part $\Sigma''$ changes sign
\begin{eqnarray}
-\Sigma''(E+is)=\Sigma''(E-is)=\pi\Delta(E)\quad s\to +0.
\end{eqnarray}
So the integral
along the branch cut is
\begin{eqnarray}
\label{cut}
I_{cut}=\int_{E_b}^{E_t}\frac{\Delta(E)e^{-iEt}dE}
{\left[E-\epsilon-\Sigma'(E)\right]^2+\pi^2{\Delta^2(E)}}.
\end{eqnarray}
Thus we have
\begin{eqnarray}
\label{ft0} g(t)=I_{cut}(t),
\end{eqnarray}
and the survival probability $p(t)$ is
\begin{eqnarray}
p(t)=|g(t)|^2.
\end{eqnarray}

In the perturbative regime
$|\Sigma'(\epsilon)|,|\Sigma''(\epsilon)|\ll
\epsilon-E_b,E_t-\epsilon$ the main contribution to the integral
(\ref{cut}) comes from the region $E\sim \epsilon$. Hence the
integral can be presented as
\begin{eqnarray}
\label{app}
I_{cut}=\int_{-\infty}^{\infty}\frac{\Delta(\epsilon)e^{-iEt}dE}
{(E-\epsilon-\Sigma'(\epsilon))^2+\pi^2{\Delta^2(\epsilon)}}
\end{eqnarray}
and easily calculated to give the well known Fermi's golden rule
(FGR)
\begin{eqnarray}
\label{fgr}
p(t)=e^{-t/\tau},
\end{eqnarray}
where $1/\tau=2\pi\Delta(\epsilon)$.

However, even in the perturbative regime, the FGR has a limited
time-domain of applicability \cite{cohen}. For large $t$  the
survival probability  is determined by the contribution to the
integral (\ref{cut}) coming from the end points. This contribution
can be evaluated even without assuming that the coupling is
perturbative. Let $\Delta(E)\sim (E-E_b)^{\beta}$ $\beta >0$) near
the band bottom. Then  for large $t$
\begin{eqnarray}
\label{b}
 I^{(b)}_{cut}\sim t^{-(\beta+1)}.
\end{eqnarray}
The similar contribution comes from the top of the band.
For the case $\beta=0$, from Eq. (\ref{sigma3}) follows that near the band
bottom
\begin{eqnarray}
\Sigma'(E)\sim \ln(E-E_b).
\end{eqnarray}
Hence in this case for large $t$
\begin{eqnarray}
\label{c}
 I^{(b)}_{cut}\sim (t\ln t)^{-1}.
\end{eqnarray}

If there can exist  poles of the propagator (\ref{exact}),  we
should add  the  residues to the integral (\ref{cut}). Thus we
obtain
\begin{eqnarray}
\label{ft}
g(t)=I_{cut}(t)+\sum_jR_j,
\end{eqnarray}
where the index $j$ enumerates all the real poles $E_j$ of the integrand, and
\begin{eqnarray}
\label{resid}
R_j=\frac{e^{-iE_jt}}{1-
\left.\frac{d\Sigma'}{dE}\right|_{E=E_j}}.
\end{eqnarray}
(In the Appendix  Eq. (\ref{ft}) is  generalized to the case of non-interacting
Fermi gas at finite temperatures.)
Notice, that the poles correspond to the energies of bound states
which can possibly occur for $E<E_b$ or $E>E_t$, and which are given
by the Equation
\begin{eqnarray}
E_j=\epsilon+\sum_{k} \frac{|V_k|^2}{E_j-\omega_k}.
\end{eqnarray}
If we take into account that normalized bound states are
\begin{eqnarray}
\left|E_j\right>=\frac{\left|d\right>+\sum_{k}
\frac{V_k}{E_j-\omega_k}\left|k\right>}{\left[1+\sum_{k}
\frac{|V_k|^2}{\left(E_j-\omega_k\right)^2}\right]^{1/2}},
\end{eqnarray}
then the residue can be easily interpreted as the amplitude of the
bound state in the initial state $\left|d\right>$, times the
evolution operator of the bound state times the amplitude of the
state $\left|d\right>$ in the bound state
\begin{eqnarray}
R_j=\left<d\right.\left|E_j\right>\left<E_j\right.\left|d\right>e^{-iE_jt}.
\end{eqnarray}
 If the propagator has one real pole at $E_1$, from Eq. (\ref{ft})
we see that the survival probability $p(t)\to |R_1|^2$ when $t\to
\infty$. If there are several poles, this equation gives Rabi
oscillations.
Notice that in perturbative regime the poles, even if
they exist, are exponentially close to the band ends, and their residues are
exponentially small.

The FGR is not valid  for small $t$ either. (From Eq. (\ref{dif}) it
is obvious that the expansion of $g(t)$ is $g(t)=1+kt^2+\dots$,
which gives quadratic decrease of the non-decay probability at small
$t$.)

Notice that Eq. (\ref{ft}) is just the well known
result \cite{mahan}
\begin{equation}
g(t)=\int_{-\infty}^{\infty}A(\omega)e^{-i\omega t}d\omega,
\end{equation}
where
\begin{equation}
\label{spectral}
A(\omega)=-\frac{1}{\pi} \text{Im}\left[g(E+is)\right]
\end{equation}
is the spectral density function. The first term in Eq. (\ref{ft})
is the contribution from the continuous spectrum, and the second
term is the contribution from the discrete states.

An indication that there is more in the GF than we have so far
discussed comes from the following fact: we could have obtained the
FGR in perturbative regime  directly from Eq. (\ref{int}), changing
exact Green function (\ref{exact}) to an approximate one, which may
be called the FGR propagator
\begin{eqnarray}
\label{fgrr}
g_{FGR}(\omega)=\frac{1}{\omega-\epsilon-\Sigma'(\epsilon)+i\pi\Delta(\epsilon)}.
\end{eqnarray}
 Thus approximated,
propagator has a simple pole $\omega=\epsilon+ \Sigma(\epsilon)$,
and the residue gives Eq. (\ref{fgr}).
Notice, that whichever approximation we use for $\Sigma(\omega)$,
the property $a(t=0)=1$ is protected, provided $\Sigma$ does not have
singularities in the upper half-plane.

These results presented above  can be illustrated by two  simple
examples.

The first example is defined by the equation
\begin{eqnarray}
\label{const} \Delta(E)=\Delta_0=\text{const}\qquad
\text{for}\;|E|\leq 1.
\end{eqnarray}
Thus we get
\begin{eqnarray}
\Sigma(\omega)=\Delta_0\log\left(\frac{\omega+1}{\omega-1}\right).
\end{eqnarray}
The Riemann surface has an infinite number of sheets. The standard
sheet is obtained by defining $\log$ as having the phase $-\pi$ just
above the real axis between $-1$ and $1$. This sheet always has  two
real poles, one for $\omega>1$, and the other for $\omega<-1$.

There are two real poles of the locator, given by the Equation
\begin{eqnarray}
\label{real}
E-\epsilon-\Delta_0\ln\left|\frac{E+1}{E-1}\right|=0.
\end{eqnarray}

In fact, in this regime from Eq. (\ref{real}) we obtain
\begin{eqnarray}
E_1=-1-e^{-\frac{1+\epsilon}{\Delta_0}},\;
E_2=1+e^{-\frac{1-\epsilon}{\Delta_0}},
\end{eqnarray}
 with the residues being equal to
\begin{eqnarray}
\label{res}
 R_1=\frac{e^{-\frac{1+\epsilon}{\Delta_0}}}{\Delta_0},\quad
R_2=\frac{e^{-\frac{1-\epsilon}{\Delta_0}}}{\Delta_0}.
\end{eqnarray}
When the locator
does not have real poles,  the  survival probability
for large time is determined by the
contribution to
the integral (\ref{cut})
coming from the end points. This contribution can
be evaluated even without assuming that the coupling is perturbative.

\begin{figure}
\includegraphics[angle=0,width=0.45\textwidth]{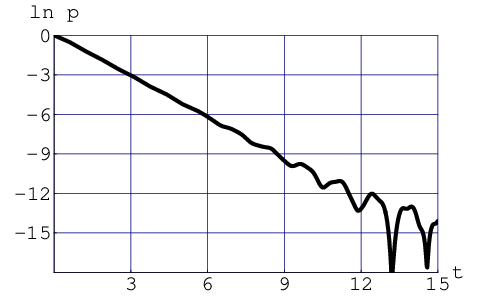}
\caption{Survival probability as a function of time  for $\Delta_0=.02$.}
\label{.02}
\end{figure}

\begin{figure}
\includegraphics[angle=0,width=0.45\textwidth]{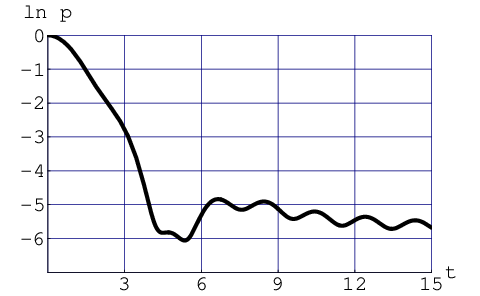}
\caption{Survival probability as a function of time for $\Delta_0=.1$.}
\label{.1}
\end{figure}

\begin{figure}
\includegraphics[angle=0,width=0.45\textwidth]{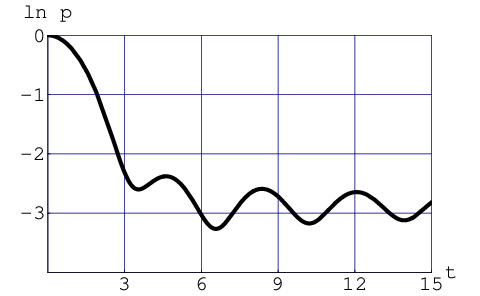}
\caption{Survival probability as a function of time for $\Delta_0=.2$.}
\label{.2}
\end{figure}

 For the sake of illustrating the results obtained above let us
presents the results of numerical calculations for the model
considered.
 The time will be measured in units of the  FGR time $\tau$
\begin{eqnarray}
1/\tau=2\pi\Delta_0.
\end{eqnarray}
For the sake of definiteness we will chose $\epsilon =-.4$. For
$\Delta_0=.02$ (see Fig. \ref{.02}) we observe  the FGR regime, say, up to
$t=9$.
For $\Delta=.1$ (see Fig. \ref{.1})  the FGR regime is seen up to $t=3$.
For $\Delta_0=.2$ (see Fig. \ref{.2})  the FGR regime is absent.
The Rabi oscillations we see already at Fig. \ref{.1} and still more vividly
at Fig. \ref{.2}.

As a second (and more physical) example
consider a site coupled to a semi-infinite lattice
\cite{longhi,visuri}. The system is described by the tight-binding
Hamiltonian
\begin{eqnarray}
\label{ham2}
&&H=-\frac{1}{2}\sum_{n=1}^{\infty}\big(\left|n\right>\left<n+1\right|
+\left|n+1\right>\left<n\right|\big)
\nonumber\\
&&+\epsilon\left|d\right>\left<d\right|-V\big(\left|d\right>\left<1\right|
+\left|1\right>\left<d\right|\big),
\end{eqnarray}
where $\left|n\right>$ is the state localized at the $n$-th site of the
lattice. The band (lattice) states are described by the Hamiltonian
\begin{eqnarray}
H_0=-\sum_k\cos k\left|k\right>\left<k\right|,
\end{eqnarray}
where $\left|k\right>=\sqrt{2N}\sum_n\sin (kn)\;\left|n\right>$.
Hence we regain Hamiltonian (\ref{ham}) with $V_k=-\sqrt{2}V\sin k$.
After simple algebra we obtain (in the upper half plain)
\begin{eqnarray}
\label{sig}
\Sigma(\omega)=\Delta_0\left(\omega-\sqrt{\omega^2-1}\right),
\end{eqnarray}
where the square root is defined as having the phase $\pi/2$ just
above the real axis between $-1$ and $1$, and $\Delta_0=2V^2$. We
immediately see that the GF for this model is a double valued
function, the branch points being $+1$ and $-1$. The poles are given
by the equation
\begin{eqnarray}
\omega_{1,2}=\frac{\epsilon(1-\Delta_0)\pm\Delta_0\sqrt{\epsilon^2-1+2\Delta_0}}{1-2\Delta_0}.
\end{eqnarray}
One sheet has real poles for $\Delta_0\geq(\epsilon^2+1)/2$. For
$\Delta_0=(\epsilon^2-1)/2$ the GF has a second order pole at
$\omega=(\epsilon^2+1)/2\epsilon$. When $\Delta_0$ increases, this
second order pole is split into two first order poles, one going
right (we assume  $\epsilon>0$) and at $\Delta_0=1/2$ becoming a
pole at the infinity. For $\Delta_0>1/2$ this pole appears for
$\omega<-1$. The second  first order pole, when $\Delta_0$ increases
initially approaches the point $\omega=1$, and at a further increase
of $\Delta_0$ moves in the opposite direction and asymptotically
goes to infinity.

\begin{figure}
\includegraphics[angle=0,width=0.45\textwidth]{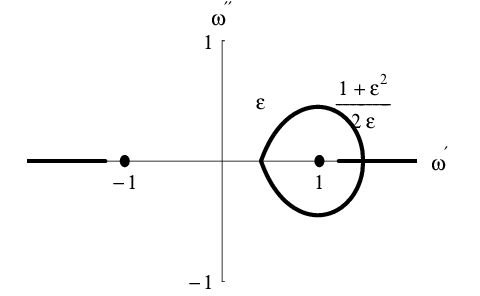}
\caption{Position of poles of the GF for the Hamiltonian
(\ref{ham2}) for different values of $\Delta$. The real poles appear
on the standard sheet, the complex poles on the second sheet.}
\label{poles}
\end{figure}

For $\Delta_0<(\epsilon^2+1)/2$  the second sheet  has two complex
poles of the first order. For $\Delta_0\ll 1$ the pole in the lower
half-plain is situated at $\epsilon -i\Delta_0\sqrt{1-\epsilon^2}$
and is just the FGR pole mentioned above. The poles position is
presented on Fig. \ref{poles}.

If we consider the case $\epsilon=0$ and $V^2/2<1$ the locator does
not have real poles, so Eq. (\ref{cut}) after substitution of the
results of Eq. (\ref{sig}) gives the amplitude of the non-decay amplitude
\cite{longhi}:
\begin{eqnarray}
g(t)=\frac{1}{2\pi}\int_{-\pi}^{\pi}dQ\exp(it\cos Q)\frac{1-\exp(-2iQ)}
{1+\alpha^2\exp(-2iQ)},
\end{eqnarray}
where  $\alpha^2=1-\Delta_0$.

\section{Analytic continuation}

For tunneling into continuum  the points $E_b$ and $E_t$ are the $\Sigma(\omega)$ (and hence the propagator (\ref{exact}))
branch points.
 Hence propagator is a
multi-valued function and it's value  in the lower half-plane
depends upon the curve
along which
we continue the function from the upper  half-plane \cite{silverman}.
The standard analytic continuation we used previously, consists of making the cut
along the straight line between the branch points and  considering
only one sheet, thus making the analytic continuation into the lower
$\omega$ half-plain by continuing $\Sigma(\omega)$ along the curves
which circumvent the right branch point clockwise and the and the
left branch point anti-clockwise.

\begin{figure}
\includegraphics[angle=0,width=0.45\textwidth]{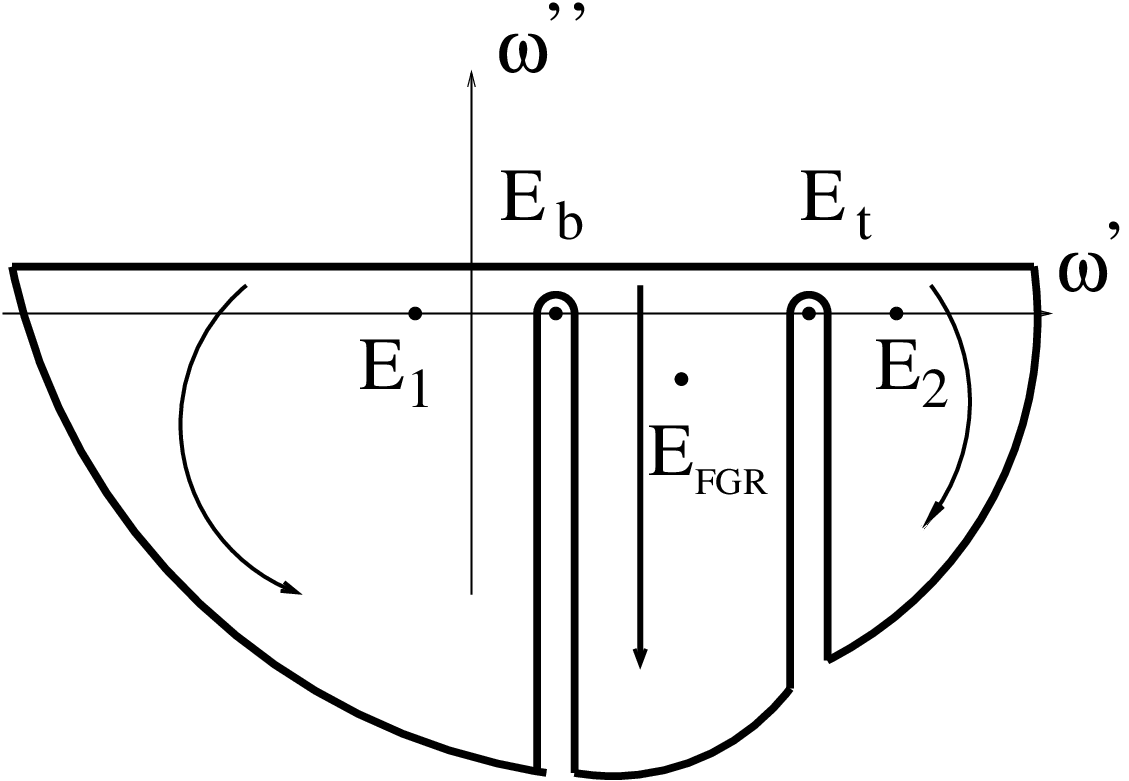}
\caption{\label{fig:cont2} Alternative way to analytically continue
the propagator into the lower half-plane. The arrows show the curves
of analytical continuation in between the cuts and outside.
 Radius of the big arc and length of the cuts go to infinity.}
 \label{altern}
\end{figure}

 On the other hand, we could use a different continuation,
making the cuts from the branch points to
infinity and continuing the function between the cuts
along the curves
passing  through the part of real axis between
$E_b$ and $E_t$, and outside as we did it previously.
This way to make analytic continuation, and hence to calculate the
integral (\ref{int}) is presented on Fig. \ref{altern}. (Of course, the value of the
integral does not depend upon the analytic continuation we use.)

(The treatment of tunneling from the discrete level into a
semi-bound continuum presented in the paper by Onley and Kumar
\cite{onley}, corresponds, in fact, to an  analytic continuation
similar in spirit, to that presented on Fig.  \ref{altern}.) Notice, that
because of the exponential decrease as a function of $t$ of the
integrand in the cut integrals appearing in this analytic
continuation, in contrast to oscillatory behavior of the real axis
cut integral, such analytic continuation is more convenient for the
numerical calculations of the large $t$ behavior of the non-decay
amplitude.

For the alternative analytic continuation, Eq.
(\ref{fgrr}) which is valid in the perturbative regime near the
point $\omega=\epsilon$ in the upper half-plane is valid in lower
half-plain also, thus giving a pole in a different sheet of the
multivalued propagator. A simple illustration of the fact that
 the complex pole  is present on the sheet of the
propagator presented on Fig. \ref{altern}, but not on that presented on Fig. \ref{contour}
is obtained in  the perturbative regime.
In the lower half-plane in the vicinity of $\omega=\epsilon$ the standard
propagator is
\begin{eqnarray}
g(\omega)=\frac{1}{\omega-\epsilon-\Sigma'(\epsilon)
-i\pi\Delta(\epsilon)},
\end{eqnarray}
and the propagator continued according to Fig. \ref{contour} is
\begin{eqnarray}
g(\omega)=\frac{1}{\omega-\epsilon-\Sigma'(\epsilon)
+i\pi\Delta(\epsilon)}.
\end{eqnarray}
Thus only the second propagator has (in a perturbative regime) a
FGR pole.

\section{Conclusions}

In this paper we solve the problem of tunneling from a discrete
level into continuum. We show, how the basic notion of the GF
formalism, like frequency dependent discrete level propagator, self
energy and spectral density, appear within the basic quantum
mechanics. We concentrate on the calculation of the  time dependent
non-decay amplitude, the stage which is typically not given proper
attention to within the GF formalism \cite{mahan}. We show that the
exponential time dependence of the non-decay probability  given by
the FGR is an approximation valid  in perturbative regime and only
for intermediate times. The large time dependence of the non-decay
probability depends crucially upon the details of the band structure
and hybridization interaction. We also look closely at those
analytic properties of the propagator in the complex $\omega$ plane,
which  often pass unnoticed.

\section{Acknowledgements}

The paper was finalized during the author's  visit to
Max-Planck-Institut fur Physik komplexer Systeme, Dresden.
The author  cordially thanks  the Institute for the hospitality extended to him during
that and all his  previous visits.

The author wishes to thank M. Katsnelson for the discussions which, in fact,
triggered this work, and R. Gol and H. Elbaz for their constant interest.

\section{Appendix}

To generalize Eq. (\ref{ft}) to the case of non-interacting
Fermi gas at finite temperatures,
let us present the Hamiltonian (\ref{ham}) using second quantization
\begin{eqnarray}
\label{ham3}
H=\sum_k\omega_kc_k^{\dagger}c_k+\epsilon d^{\dagger}d
+\sum_k\left(V_kc_k^{\dagger}d+h.c.\right),
\end{eqnarray}
where $c_k^{\dagger}(c_k)$ and $d^{\dagger}(d)$
are creation (annihilation) operators of band states and
discrete  state respectively. The tunneling  of either the
electron or the hole from
the discrete level into continuum is described by Green's functions
$G^{>}$ and $G^{<}$ respectively
\cite{mahan}
\begin{eqnarray}
G^{>}(t)&=&\left<d(t)d^{\dagger}(0)\right>\nonumber\\
G^{<}(t)&=&\left<d^{\dagger}(t)d(0)\right>.
\end{eqnarray}
where the averaging is with respect to the grand
canonical ensemble, and $d(t)$ or $d^{\dagger}(t)$ is the
annihilation or the creation operator in Heisenberg representation.
Both Green's functions are simply connected with the spectral density function
\cite{mahan}
\begin{eqnarray}
G^{>}(\omega)&=&[1-n_F(\omega)]A(\omega)\nonumber\\
G^{<}&=&n_F(\omega)A(\omega),
\end{eqnarray}
where
$n_F(\omega)=\left(e^{\beta(\omega-\mu)}+1\right)^{-1}$
is the Fermi distribution function ($\mu$ is the chemical potential and $\beta$
is the inverse temperature).
Thus we obtain
\begin{eqnarray}
\label{ft3}
G^{>}(t)&=&\int_{E_b}^{E_t}\frac{\Delta(E)[1-n_F(E)]e^{-iEt}dE}
{\left[E-\epsilon-\Sigma'(E)\right]^2+\pi^2{\Delta^2(E)}}\nonumber\\
&+&\sum_j[1-n_F(E_j)]R_j\nonumber\\
G^{<}(t)&=&\int_{E_b}^{E_t}\frac{\Delta(E)n_F(E)e^{-iEt}dE}
{\left[E-\epsilon-\Sigma'(E)\right]^2+\pi^2{\Delta^2(E)}}\nonumber\\
&+&\sum_jn_F(E_j)R_j.
\end{eqnarray}

\end{document}